# Surface Structure of Organoclays

Hendrik Heinz and Ulrich W. Suter*

* Prof. Dr. Ulrich Werner Suter, Dr. Hendrik Heinz

   Institute of Polymers, Department of Materials

   ETH Zürich

   CH-8092 Zürich

   Switzerland

   Fax: +41 1 632 15 92

   E-mail: uwsuter@eth.ch

** We thank Prof. Dr. Wolfgang Paul, Department of Physics, University of Mainz, Germany, and Prof. Dr. Andrey Milchev, Department of Physics and Astronomy, University of Athens, Georgia, USA, for helpful discussions. We acknowledge support from the ETH Zürich, the Swiss National Science Foundation, and the Studienstiftung des Deutschen Volkes.

# Abstract

The properties of organically modified clay minerals determine essentially the quality of polymer-clay nanocomposites. We investigate alkylammonium-micas with different alkyl chain length by molecular dynamics simulation as homogeneous mixtures and separated island structures on the mica surface (80 % alkali exchange, 20 % of the alkali ions remain), including a detailed model of the Si → Al ⋯ K charge defects on the surface of phyllosilicates. By comparison with experiment, we find that long chains ($\geq C_{18}$) lead to a mixed phase of alkali ions and organic ions on the surface, while chains of a medium length ($\sim C_{12}$) give rise to phase separation. Very short chains, by thermodynamical arguments, prefer again a homogeneously mixed surface. We explain this interesting observation with a basic free energy model and present a diagram of the phase behaviour as a function of surface saturation with alkyl chains and the chain length. Besides, an order-disorder transition of the tethered alkyl chains is found on heating when ~20 % of the torsional angles along the hydrocarbon backbones are *gauche*. At higher temperature, also rearrangements of the ammonium headgroups on the surface are possible, which lead to metastable structures after quenching.

Mica with amphiphilic cations is of extensive technological interest. The mechanical stability of various plastics and rubbers, as well as barrier and optical properties in thin films can be substantially improved by addition of organically modified mica.[1] Besides, mica surfaces have proved to be an excellent model system for surface phenomena because the mineral can be cleaved into regular surfaces that extend over μm. The inorganic-organic interfaces have been the subject of numerous experimental studies[2-8] and theoretical investigations of related systems were performed at coarse-grained[9, 10] and atomistic levels.[11-14] Accurate molecular dynamics simulation at atomistic level supports the interpretation of experiments, e. g., XRD and solid-state NMR data.[7, 15] With our recent simulations,[13, 14] the inclination angles of the alkyl chains and basal-plane spacings of the filler particles are reproduced in quantitative agreement with experiments, and predictions of the interface structure as well as conformational analyses of the hydrocarbon chains are possible. Here, we report results on monolayer phases on mica platelets with alkali ions and surfactant ions of different length, at different temperatures, and give insight into the occuring phase transitions.[6, 7, 13, 14]

We consider dry mica surfaces where 80 % of the alkali ions, i. e., most commonly lithium in the referenced experimental studies, have been exchanged by organic ammonium ions of different length (Scheme 1) so that 20 % of the alkali ions remain. This is a technologically realistic situation in view of the relatively difficult quantitative replacement of all interlayer cations (without intercalation). We investigate two borderline cases. The first simulated structure is a homogeneous mixture of surfactant ions and alkali ions on the mica surface. The second simulated structure refers to "phase-separation", i. e., segregation of cations at the surface. In that case, we model the surfactant islands on mica where the alkali exchange is quantitative. In both cases, we employ a mica model containing the upper half of 5×3×1 unit cells with realistic atomic charges.[13, 16] For the homogeneous phase, twelve dialkyldimethylammonium ions (Scheme 1) and three potassium ions are attached, accounting for 80 % cation exchange. For the simulation of "islands", 15 dialkyldimethylammonium ions are added to the surface, corresponding to 100 % ion exchange. All structures are periodic in the *xy* plane and open in the *z* direction.[17] At the outset, the plain, single-coated (mica-alkyl) structures are subjected to NVT dynamics (time step = 1 fs) at a given temperature and tilt angles of the alkyl chains are determined. Thereafter, duplicate assemblies (mica-alkyl-alkyl-mica) are constructed and equilibrated again (criterion: no more structural changes, >400 ps). 100 snapshots are subsequently taken at intervals of 1 ps to compute the system's properties, such as basal-plane spacings and conformations of the alkyl chains. We conduct our calculations with the extended consistent force field 91, which is accurate in modeling organically modified silicates and was previously described together with other simulation details,[13, 18, 19] using the Discover program from MSI.[20] We note that the NpT ensemble is not required for the simulation (instead of NVT) because an added atmospheric pressure of ~0.1 MPa does practically not affect the geometry of our condensed-matter system. The reason are the elastic moduli of roughly 1 GPa in *z*-direction and >15 GPa in *x*- and *y*-direction.[21] Accordingly, differences in the simulation are negligible (see also the detailed analysis of the pressure and pressure profiles of our system in ref. [14]).

Since we use a larger simulation box than in our previous study,[13, 14] we consider the distribution of Al-defects on the mica surface in a more elaborate way (Figure 1).[22-24] For a ratio Al/Si = 1/3 on the surface of natural mica, we see from Figure 1a that (1) 60 % of the surface Si atoms are connected via oxygen atoms to 1 Al atom and 2 Si atoms, (2) 20 % of the Si atoms are connected via oxygen to 0 Al atoms and 3 Si atoms, and (3) 20 % of the Si atoms are connected via oxygen to 2 Al atoms and 1 Si atom. Also, Al-O-Al contacts do not exist (Figure 1). With these statistical criteria, we obtain a more realistic distribution of Al-defect sites than in the regular arrangements.[13] As can be shown by Molecular Mechanics, the alkali ions and ammonium headgroups preferably reside over cavities containing two or three Al-defects along their boundary (Figure 1b). This is a consequence of electrostatic forces, whereby precedence is given to the alkali ions, which have a greater charge density per volume relative to the more bulky tetraalkylammonium group. The distribution of the three $Li^+$ ions in the simulation box for a homogeneous phase is chosen uniform (Figure 1b). Besides, we note that our simulation box with ~$10^1$ alkyl chains is small compared to the $10^4$ to $10^6$ alkyl chains on a µm-sized mica flake.[5]

Let us now turn to the simulation results after more than 400 ps. The lattices of the ammonium headgroups are relatively stable for all cations from $2C_{12}$ to $2C_{18}$ in all cases considered, but rearrangements across surface cavities are, in principle, always possible and have been observed in the course of the simulations in several instances. At elevated temperatures, some ammonium ions change to another cavity, especially in the homogeneous structures where rearrangements are geometrically easier to achieve. A representative example of ammonium headgroup positions for $2C_{18}$ ions in the homogeneous phase is displayed in Figure 1b.

It is, however, not possible to decide from the simulations, which limiting structure (homogeneous or segregated phases) is preferred, because we cannot simulate a large system with two separate phases. We will, therefore, consult the experimental data.[7] Phase-separated and homogeneous structures can be distinguished by their different basal-plane spacings:[13] Phase-separated structures always have a higher basal-plane spacing than the corresponding homogeneous structure because the orientation of the surfactant chains is nearer to the surface normal. In Figure 2, the computed basal-plane spacings for one-phase and two-phase surfaces are

 

compared with the experimental values[7] at low and high temperatures. Both graphs indicate that the $2C_{18}$ chains are homogeneous with the alkali ions on the surface. With decreasing chain length, a trend towards phase segregation between organic chains and alkali ions is apparent. The $2C_{12}$ ions are fully organized into islands. The result is supported by the close match between experimental and computed basal-plane spacings for $2C_{18}$-mica and $2C_{12}$-mica (Table 1).

Also, the thermal behaviour reported by Osman et al.[7] for mica covered with $2C_{18}$, $2C_{16}$, $2C_{14}$, and $2C_{12}$ chains can be explained:

(1) For $2C_{18}$ chains, only one sharp phase transition is observed with DSC (Differential Scanning Calorimetry) on heating, which is reversible upon subsequent cooling and reheating.[7] It is in accordance with a homogeneously mixed surface structure where the tethered alkyl chains undergo a reversible partial melting without significant rearrangements (see Figure 3a/b). We count the number of *gauche*-arrangements from the end of one $C_n$ chain via the nitrogen atom to the end of the second $C_n$ chain. $2C_{18}$ chains, which have generically ~4 *gauche*-torsions near the ammonium head group, possess then roughly 3.3 *gauche*-arrangements per $C_{18}$ backbone and a significant tilt angle of 29° (Table 1) below the order-disorder transition at 55 °C.[7] In the partially molten state, no tilt angle can be specified.

(2) For $2C_{14}$ chains, one sharp DSC signal of relatively small enthalpy at ~35 °C is observed upon heating, which produces a metastable phase.[7] This phase exhibits subsequent transitions on cooling and reheating at the same temperature, but they are weaker and indistinct.[7] These facts are in agreement with substantial accumulation of the $C_{14}$ ions into island-like structures and some mixed domains on the mica surface as conluded from Figure 2. There is a reduced possibility for conformational disorder (significantly smaller area of the DSC peak and less pronounced changes in the IR spectrum compared to $2C_{18}$ chains during the transition[7]). In the sharp transition, the chain backbones are "partially melting" with some rearrangements of the ammonium headgroups occuring simultaneously. Subsequent cooling and reheating reveals the instantaneous freezing and re-melting of the obtained metastable phase with rearranged headgroup positions (less well-defined transition). The original structure is formed after several hours at room temperature because reverse rearrangements to reorganize the displaced headgroups are slow at ambient temperature.[6, 13]

(3) For 2$C_{12}$ chains, neither a phase transition in the IR spectrum nor a pronounced change in $^{13}$C-NMR signals at higher temperature is found (DSC results not shown),[7] and the basal-plane spacing is also not significantly changing from 60 °C to 20 °C (not shown at –20 °C).[7] The chains seem to be tightly packed in the islands and too short for extensive conformational disorder (Figure 3c/d). The number of *gauche*-incidences is only 1.0 to 1.3 per $C_{12}$ backbone (Table 1). Therefore, no order-disorder transition is possible (Figure 3 c/d) in contrast to the higher homologues.

In the following, we want to rationalize the phase behaviour of the alkyl chains on mica. We found previously[13] a gradual transition from separated phases towards a homogeneous phase upon increasing the saturation of the mica surface with alkyl chains. In this report, we illustrated that for a given degree of alkali exchange the length of the alkyl chains is also an important factor: long chains prefer a homogeneous phase, chains of a medium length lead to island formation, and, as we surmise from the alkali-like nature of the headgroups, very short chains might prefer a single-phase system again.

Which arrangement on the mica surface appears, provided the exchange reaction is driven to equilibrium, is determined by thermodynamics. Between the two borderline cases, the structure

7 of 18    mix_alkyl_submit_AngewCh_revised6

with the lowest free energy is predominantly formed, depending on their difference in free energy, $\Delta A_{1\to 2}$ (see ref. [25] for a different example). The difference in free energy to create a 2-phase system from the 1-phase system, while the stoichiometry remains constant, $\Delta A_{1\to 2}$, is given at a certain temperature as

$$\Delta A_{1\to 2} = \Delta E_{1\to 2} - T\Delta S_{1\to 2}. \tag{1}$$

With respect to our system, we can interpret the quantities in Eq. (1) as follows: $\Delta E_{1\to 2}$ is equal to the difference in average van-der Waals energy between the chain backbones. For very low surface coverage, when the chains could not at all or scarcely interact in a 1-phase system, $\Delta E_{1\to 2}$ is negative and proportional to chain length. For any surface coverage higher than that, the chains optimize their dispersive interactions by an appropriate inclination angle in both the 1-phase and the 2-phase system and $\Delta E_{1\to 2}$ approaches zero, independent of chain length. However, a small negative value remains because the dispersion interactions near the ammonium headgroups in the islands of the 2-phase structure are better. The entropy term $\Delta S_{1\to 2}$ accounts for changes in the configurational space of our system. As indicated in the phase transitions at changing temperature, the configurational space concerns primarily order vs disorder in the chain backbones and, secondarily, order vs disorder in the headgroup arrangement on the surface. In a 2-phase system, the chain disorder is substantially reduced because the conformational freedom in the islands is far more restrained than in the homogeneous structure. Accordingly, $\Delta S_{1\to 2}$ has a negative value. Since the difference in conformational freedom is more pronounced the longer the chains are, $\Delta S_{1\to 2}$ increases substantially its negative value with chain length. Another effect that guarantees always a small negative value for $\Delta S_{1\to 2}$ is the reduction of headgroup disorder in a 2-phase system relative to the alkali ions. This value is estimated to be $\Delta S_{head} = -0.5R$ for 80 % alkali exchange,[26] although it cannot be decoupled from the conformational entropy for longer chains. We may, however, approximately write

$$\Delta A_{1\to 2} = \Delta E_{vdW} - T(\Delta S_{head} + \Delta S_{chain}). \qquad (2)$$

In consequence: (1) If the degree of ion exchange is so low that the chains cannot feel dispersive interactions with each other in a 1-phase system even with extreme tilt, $\Delta E_{vdW}$ is strongly negative and overpowers the other contributions; thus, a 2-phase system is formed. (2) If the chain length in such a system is increased such that $\Delta S_{chain}$ becomes highly negative, $\Delta A_{1\to 2}$ will be positive and a 1-phase system is preferred. (3) If the chain length is reduced towards zero, $\Delta E_{vdW}$ will be very small, $\Delta S_{chain}$ disappears, and the negative term $\Delta S_{head}$ will lead to a slightly positive $\Delta A_{1\to 2}$ of $+0.5RT$. This difference in free energy is small so that intermediate arrangements are possible, however, with the tendency towards a one-phase system. These considerations lead to a kind of "phase diagram" of surface saturation vs chain length for alkylammonium-modified clay surfaces (Figure 4).

**Legends to the Figures and Schemes**

**Scheme 1.** The dialkyldimethylammonium ions used to modify the mica surface, and their abbreviations.

**Figure 1.** a) Si microenvironments on the surface of tetrahedral layers in phyllosilicates as a function of Al substitution, as observed by $^{29}$Si-NMR.[22-24] b) Application to our mica model. The proper statistical connectivity around Si (yellow) and Al (blue), locations of the alkali ions (violet pellets), and representative positions of the headgroup N atoms (green pellets) of $2C_{18}$ ions on the mica surface are shown with 80 % lithium exchange (oxygen red).

**Figure 2.** Computed basal-plane spacings for the two limiting cases of one homogeneous phase and two phases of dialkylammonium ions (80 %) and alkali ions (20 %) on mica. The experimental values are also shown and indicate the real structure. The overall structure is not dependent on the temperature.

**Figure 3.** Snapshots of the homogeneous $2C_{18} - K^+$ phase (80 % alkylammonium, 20 % alkali ions) on mica after 400 ps of molecular dynamics a) at 0 °C and b) at 80 °C. A partial "melting" of the tethered chains is visible. Snapshots of $C_{12}$ ions (as found in the islands) on mica c) at –20 °C and d) at 60 °C. The absence of a significant change and the almost perpendicular orientation of the alkyl chains are apparent.

**Figure 4.** Approximate sketch of a "phase diagram" for alkylammonium-mica, indicating the type of system found for various degrees of surface coverage (see ref. [13]) and chain length.

**Table 1.** Computed characteristics for 2C$_{18}$ on mica where 20 % of the alkali ions remain and for 2C$_{12}$ ions on mica at 100 % ion exchange (corresponding to "islands" in the 80 % coverage systems): basal-plane spacings (in nm), average number of *gauche*-incidences in the alkyl chains, and their tilt angles (in deg).

|  |  | Basal plane spacing | | No. of *gauche*-arrangements | | | Tilt angle |
| --- | --- | --- | --- | --- | --- | --- | --- |
|  |  | Exptl[a] | MD-simulation | Total | backbones | Max. poss. |  |
| 2C$_{18}$ | 0 °C | 4.7–4.8 | 4.76 | 9.6 | ~5.6 | 34 | 29 ±5[b] |
|  | 80 °C | 5.02 | 4.96 | 11.5 | ~7.5 | 34 | —[c] |
| 2C$_{12}$ | −20 °C | – | 4.14 | 5.9 | ~1.9 | 22 | 5 ±4 |
|  | 60 °C | 4.33 | 4.22 | 6.7 | ~2.7 | 22 | 5 ±4 |

[a] Ref. 7. [b] The orientation in the layer with interspersed alkali ions is not very strict. [c] Too disordered.

**Suggestion for the Table of Contents**

While long and very short alkyl chains in alkylammonium-mica form a homogeneous mixture with remaining alkali ions on the mica surface, chains of medium length (~$C_{12}$) prefer island structures. We point out this observation in detail, and give a thermodynamic explanation. The surface structure is essential for the design of organically modified clay components in polymer-clay nanocomposites.

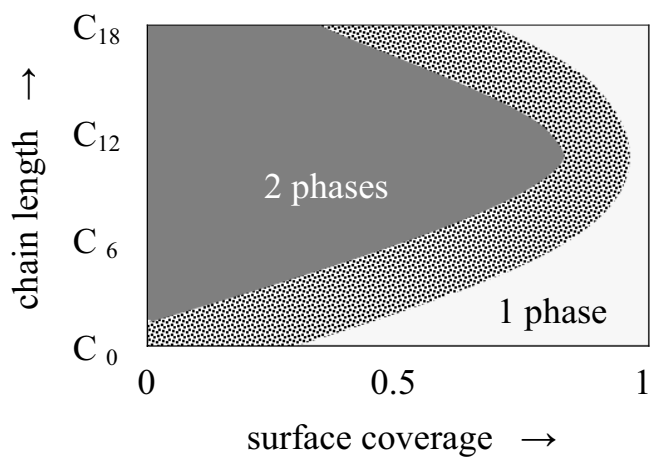

**Keywords**

organic-inorganic interfaces, molecular dynamics, conformational analysis, phase diagram, free energy model, mica

**Graphical Material**

Equation (1)  $\Delta A_{1\to 2} = \Delta E_{1\to 2} - T\Delta S_{1\to 2}$

Equation (2)  $\Delta A_{1\to 2} = \Delta E_{vdW} - T(\Delta S_{head} + \Delta S_{chain})$

**Scheme 1**

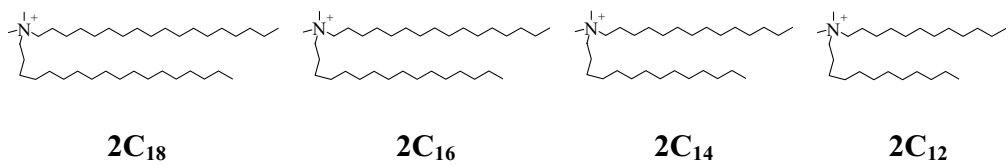

    **2C$_{18}$**        **2C$_{16}$**        **2C$_{14}$**        **2C$_{12}$**

**Figure 1**

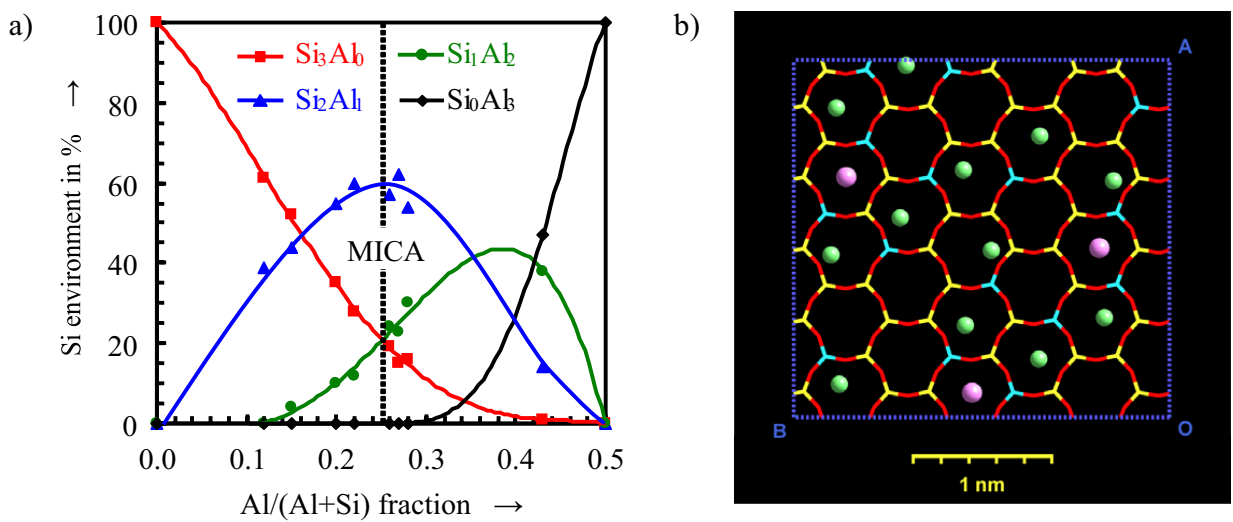

**Figure 2**

a)

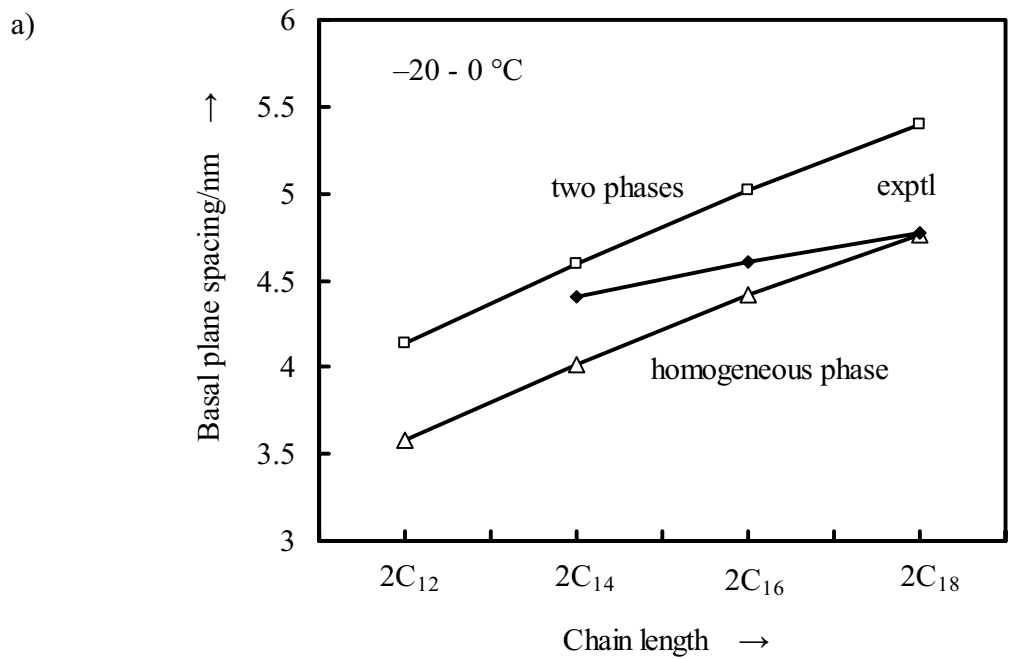

b)

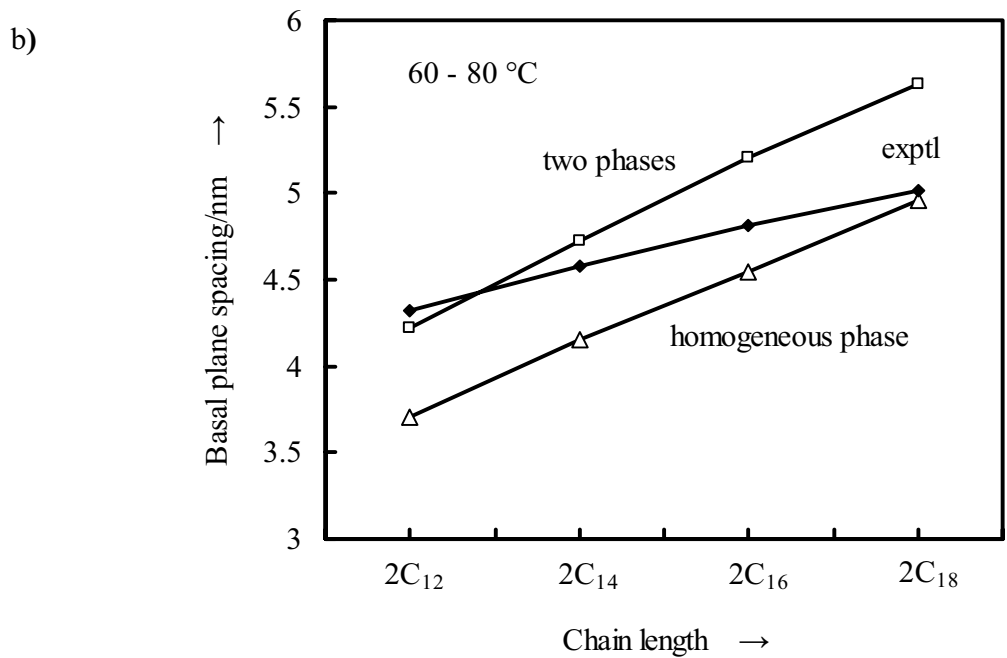

**Figure 3**

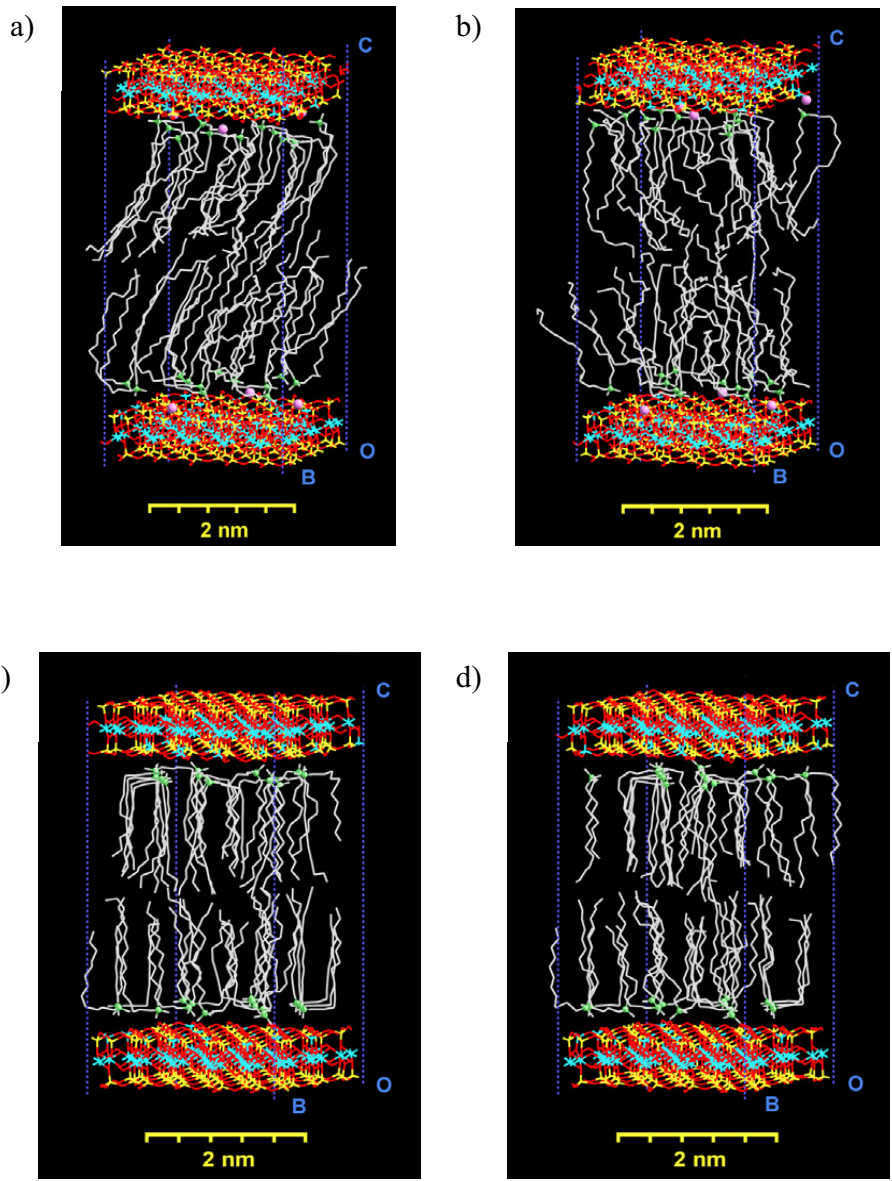

 

**Figure 4**

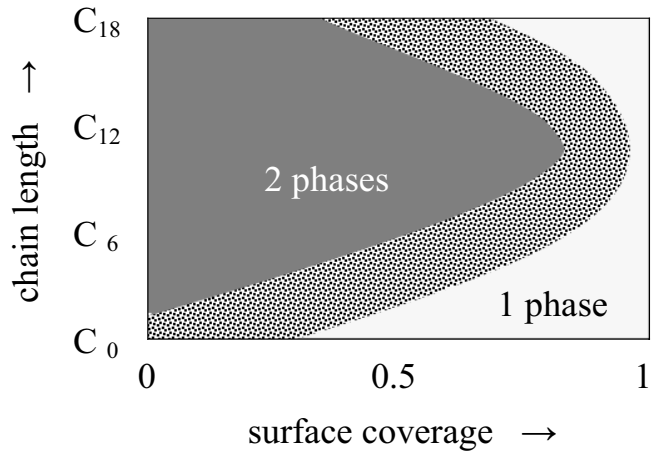